\documentclass[10pt,twocolumn]{article}
\usepackage{latex8}
\usepackage[english]{babel}
\usepackage{epsfig}
\usepackage{tabularx,supertabular}
\usepackage{latexsym}
\usepackage{amsmath}
\usepackage{amssymb}
\def\nmr{\hbox{$N$\kern-1pt MR}}

{\bf}{\rm}
{\bf}{\rm}
{\bf}{\rm}
{\bf}{\rm}
\def\qed{\relax\ifmmode\hskip2em \Box\else\unskip\nobreak\hskip1em $\Box$\fi}
\def\coldot{,} 
 {\catcode`\.=\active\gdef.{$\egroup\setbox2=\hbox to \dimen0 \bgroup$\coldot}}
\def\rightdots#1{%
  \setbox0=\hbox{$1$}\dimen0=#1\wd0
  \setbox0=\hbox{$\coldot$}\advance\dimen0\wd0
  \setbox2=\hbox to \dimen0 {}%
  \setbox0=\hbox\bgroup\mathcode`\.="8000 $}
\def\endrightdots{$\hfil\egroup\box0\box2}

\newcolumntype{d}[1]{>{\hfill\rightdots{#1}}r<{\endrightdots}}
\newcolumntype{D}[1]{>{\hfill\rightdots{#1}}X<{\endrightdots}}
\newcolumntype{C}{>{\centering\arraybackslash}X}
\newcolumntype{L}{X}
\newcolumntype{R}{>{\hfill}X}
\newlength{\widthplusfourty}
\newcommand{\ReLinda}{\hbox{$\mathcal R\!\raise2pt\hbox{$\varepsilon$}\!\hbox{$\mathcal L$inda}$}}

\begin{document}
\title{Flexible Development of Dependability Services: 
An Experience Derived from Energy Automation Systems}
\date{}
%
\author{%
  Vincenzo De Florio\\
    Katholieke Universiteit Leuven\\
      Electrical Engineering Department, ELEN-ACCA Division,\\
      Kasteelpark Arenberg 10, B-3001 Leuven-Heverlee, Belgium\\
      E-mail: deflorio@esat.kuleuven.ac.be\\
    \and
  Susanna Donatelli\\
      Dip. di Informatica, Universit\`a di Torino,\\
      C.so Svizzera 185, I-10149 Torino, Italy\\
    \and
  Giovanna Dondossola\\
      CESI, Automation Business Unit\\
      Via Rubattino 54, I-20134 Milan, Italy%
}
\maketitle
%
\begin{abstract}
This paper describes a novel approach for the flexible development 
of dependable automation services applied to a case study taken from 
requirements of energy automation systems. It shows first how the use of a 
custom compositional recovery language can be exploited to achieve a flexible and 
dependable functionality in software. Then it is shown how modeling 
techniques based on Petri nets can be used to assess the properties that different 
configurations of the addressed service can achieve.
\end{abstract}
\Section{Introduction}
The target application domain, from which the case study of the present contribution 
has been derived, concerns the automation of the Electric Power System (EPS) in 
charge of producing, transporting and distributing electricity. An EPS is characterised 
by a complex network topology whose nodes of interconnection are High Voltage 
Substations, Primary and Secondary Substations connected by high and medium 
voltage lines.
Energy automation systems have a hierarchical organisation: they perform mission-critical
automation functions (such as monitoring, command/control and protection) 
requiring different degrees of dependability, depending on their criticality degree 
(influence and propagation of faults affecting the functions, possibility and cost of 
confinement). The availability required by a function depends on its positioning 
within the hierarchical structure of the automation system: the highest is the function, 
the highest degree of availability is required; for what concerns credibility  (integrity 
and security), the lowest is the position of the function, the highest is the required 
credibility, i.e. the closest to the field, the most serious the consequences. 
Dependability needs of automation systems in the electric power domain have been 
traditionally addressed by realising custom hardware-based devices, which are 
capable of guaranteeing high availability and integrity to the automation function.
The evolution of the future generation systems, belonging to different automation 
levels, requires parallel and distributed implementations on a variety of scaleable high 
performance and fault tolerant architectures, capable to answer to the increasing 
demand for performance and dependability coming from the application field, while 
preserving the previous investments~\cite{DeBo98}.
An emerging challenge in the development of dependable systems is represented by 
the availability of generic software-based fault tolerance capabilities, which are 
portable on COTS components and easily adaptable to different configurations of the 
same application, as well as to different applications and domains.
In the ESPRIT project TIRAN  this challenge has been tackled by developing a 
flexible approach which has been ported and demonstrated on different COTS 
platforms \cite{BDDC99b}. This paper mainly describes how a component of the approach, 
the language Ariel, may be used to provide a specific and robust fault tolerance 
functionality, i.e. the Redundant Watchdog, satisfying dependability requirements 
typical of a wide class of applications in the energy domain. The behaviour of 
alternative Ariel implementations of the Redundant Watchdog is then modelled with 
Stochastic Petri Nets and some preliminary results from the evaluation of models are 
given.
The structure of the paper is as follows. First, the application requirements for a 
Watchdog mechanisms and the characteristics of  a Redundant Watchdog 
functionality are introduced in Section 2. Next, in Section 3, the Ariel language is 
introduced and specific differences with respect to similar approaches are
pointed out.
Sections 4 and 5 
respectively explain how the Ariel language can be used to realise the Redundant 
Watchdog and provide potential users with some preliminary results from the 
evaluation of Stochastic Petri Nets models. Finally, Section 6 concludes this 
contribution summarising the current achievements and the future extensions.

\Section{An Industrial Problem: The Redundant Watchdog Requirements}
Technicians of energy automation systems typically express requirements in textual 
form, capturing the main dependability needs of the applications. Considering the 
Primary Substation Automation System a list of dependability requirements has been 
collected and addressed in the TIRAN project~\cite{TIRAN:D1.1}.
In the present paper we will focus on two of those  Application Requirements 
(referred as AR1 and AR2 below) that lead to the need of an enhanced watchdog 
mechanism. They are formulated as follows:
\begin{description}
\item[AR1]: ``If an erroneous situation can not be recovered according to required mode and within 
given time constraints, then a mechanism for the auto-exclusion of the system should be 
provided which, if not reset before the expiration of a pre-fixed time-out, disconnects the system 
from the plant, leaving the plant in an acceptable state, forcing the output to assume a pre-
defined secure configuration, providing appropriate signalling to the operator and to the 
remote systems (as automation system failures should not affect the plant).''
\item[AR2]: ``The auto-exclusion should guarantee a high availability, integrity and security - e.g. by 
a redundant and periodically tested mechanism, with auto-diagnostics.''
\end{description}

The auto-exclusion functionality (as required by AR1 and AR2) has been traditionally 
supported by the so-called plant's watchdog (plantWD) mechanism, a dedicated 
hardware device with high integrity and availability degrees. In most cases the 
plantWD mechanism is used as an ultimate action of a fault tolerant strategy to detect 
un-recovered processing errors and to avoid their propagation. Errors are typically 
run-time violations occurring during the execution of an application process due to, 
e.g., a process that has crashed or is slowed down.

The watchdog mechanism (WD) cyclically sets a timer requiring an application 
process to explicitly reset it by sending an ``I am alive'' message before it reaches its 
deadline. If, for any reason, the application process is not able to send the message, 
the watchdog raises an error condition that has to be treated by some entity in some 
way. Depending on the global fault tolerance strategy adopted our plantWD is set to 
count either the double of or the same time of the basic application cycle.

In support of the migration to flexible software dependability
services
running on COTS platforms, the goal of developing a robust,
software-based
WD mechanism has been addressed by the TIRAN Project. A watchdog
basic tool
has been implemented characterised by the following Watchdog
Requirements
(WR) and Properties (WP):
\begin{description}
\item[WR1]: ``The WD has to survive at system reboot or reset, i.e. the memory it allocates for its 
counter is not to be cleared.''
\item[WR2]: ``In a distributed software architecture the application node's signals have to be put in a 
logic AND to actually signal the WD, i.e. the WD effectively stops to countdown only if on each 
node the execution has terminated correctly.''
\item[WR3]: ``The WD has to survive at node failures, i.e. whatever node faults the WD mechanism 
should be not compromised.''
\item[WR4]: In order to guarantee correct operation of the WD mechanism, it is mandatory that the 
WD task is running at a higher priority than the tasks (that run on the same node) it supervises. 
WD tasks supervising tasks on other nodes must have appropriate priority to ensure proper 
operation. It is the responsibility of the application writer to ensure correct partitioning and 
priority allocation.
\end{description}

\begin{description}
\item[WP1]: The watchdog task can be placed either on the same node where the application tasks 
run on or on a different node.

\item[WP2]: Placing the watchdog task on the same node where the application task runs on 
minimises overhead and detection latency.
\item[WP3]: Placing the watchdog task on a different node with respect to the application node 
lowers the probability of a common failure for both application and watchdog task that would 
go undetected.
\item[WP4]: Detection latency is under the control of the application writer. The higher the frequency 
of sending ``I'm alive'' messages, the lower the detection latency.
\item[WP5]: Overhead is under the control of the application writer. The lower the frequency of 
sending ``I'm alive'' messages, the lower the overhead paid by the application task and the 
communication system.
\item[WP6]: WD is just one task which receives system clock ticks and application ``I'm alive'' 
messages. Both types of messages are received through interprocess communication and are 
asynchronous to WD task.
\item[WP7]: Being the WD in a distributed software architecture it is able to receive multiple signals 
and to apply a logical operation on them (i.e. in the case of the logical operation AND required 
by WR1 the WD will fire if at least one node does not produce its signal).
\end{description}
In section 4 it will be shown how the requirement WR3 above may be fulfilled by 
instantiating more WD mechanisms and by applying different voting mechanisms to 
their firings. Such Redundant WatchDog (RWD) mechanism is characterised by the 
following design properties:
\begin{description}
\item[RWP1]: Processing errors affecting WD replicas can be detected and recovered transparently 
by the RWD
\item[RWP2]: The number of WD replicas and the voting mechanism chosen determine a different 
improvement of the RWD dependability: e.g. Nreplica=3, allows a 2-out-of-3 voting (which can 
correct up to 1 fault); the selection of the suitable Nreplica and voting is a compromise among 
dependability and performance overhead, left to the application writer's experience.
\item[RWP3]: WD replicas can be placed all on the same node. This minimises overhead and 
detection latency but does not increase the RWD dependability.
\item[RWP4]: WD replicas can be placed on different nodes. This minimises the chance of a common 
failure affecting each WD replica.
\end{description}

In section 5 the analysis of Stochastic Petri Nets models of alternative RWD policies 
will prove the properties WP2 and RWP2 above. The comparison of the computed 
performance measures provides asensitivity indication of the requirement AR2 to the 
number of WD replicas.

\begin{figure*}[tpb]

\centerline{\includegraphics[width=.8\textwidth]{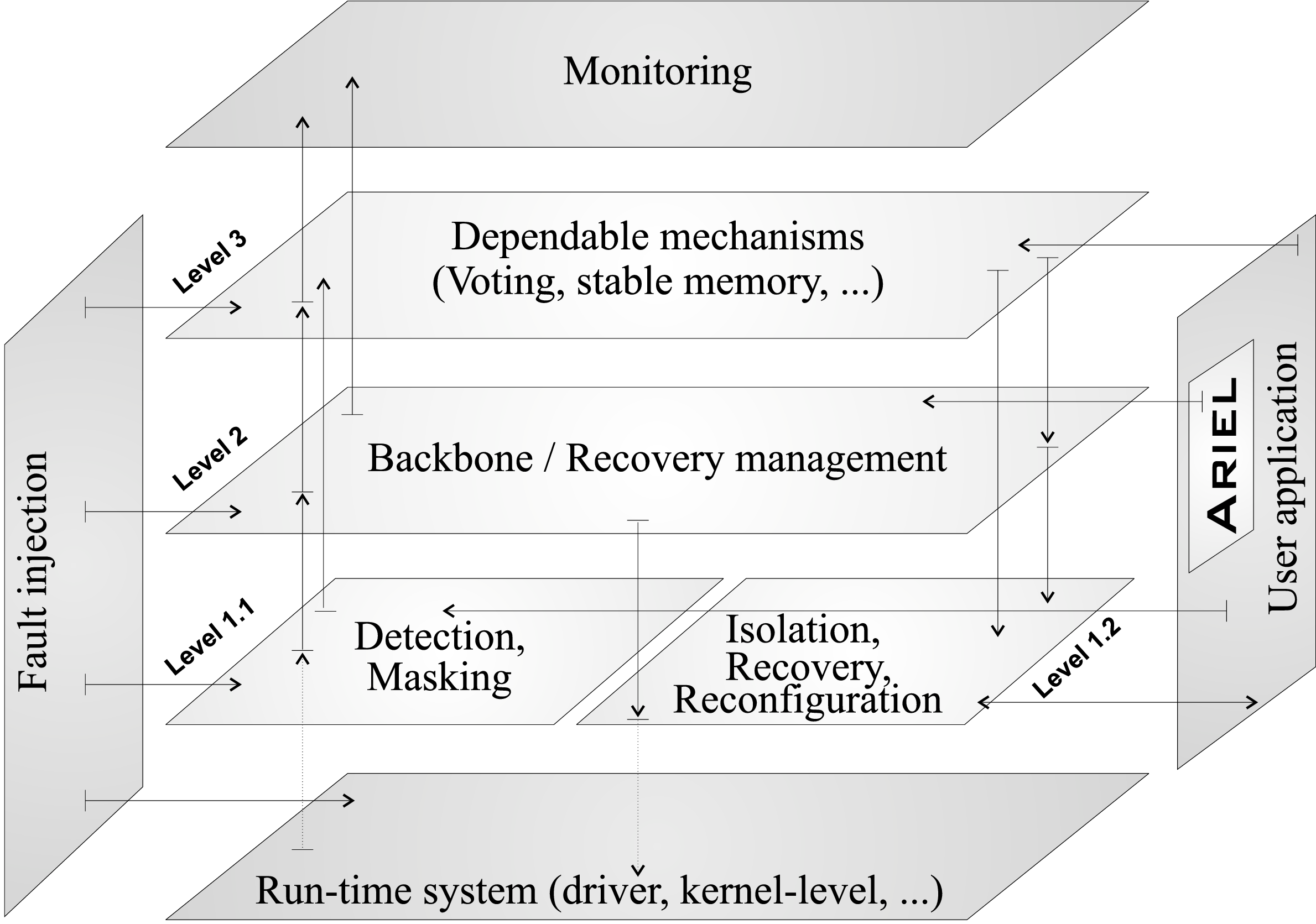}}

\caption{The TIRAN architecture and its main components.}\label{f:tiran}
\end{figure*}

\Section{The Ariel Language}
Ariel~\cite{DF00} is a recovery, configuration and coordination language that has been 
defined inside the TIRAN project. By recovery language (RL) we mean a linguistic 
framework for the expression of the error recovery aspects of a distributed 
application~\cite{DF00,DeDL99a}. According to this approach, beside the service 
language, i.e., the programming language addressing the functional design concerns, 
a special-purpose linguistic structure (the RL) is available to address error recovery 
and reconfiguration in an attempt to minimize non-functional code intrusion and 
hence to improve the separation between the functional and the fault-tolerance design 
concerns. To some extent this allows to decompose the design process into two 
distinct phases, thus  providing a way to control the design complexity and to reduce 
coding times. RL programs are executed either asynchronously with the user 
application, when an error detection tool from those in a custom library signals that an 
entity has been found in error, or synchronously, when the application itself declares 
that has entered an erroneous state (via instrumented assertions or self-checking). 
Examples of entities are: processing nodes, tasks, group of tasks (called ``logicals'').
In the RL prototyped in Ariel error recovery is specified in terms of guarded actions: 
actions specify recovery activity on entities of an application and pre-conditions query 
the current state of those entities. The state of the entities, as it appears to the 
detection tools, is sent to a middleware entity called Backbone (BB), which arranges it 
into the form of a system-wide database.
The execution of the user-specified recovery actions is done via a fixed scheme. 
Together with the application, two special-purpose tasks are running: the BB task and 
a ``recovery application'' task. As soon as an error is detected, a notification describing 
that event is sent to the BB that stores the notification and starts the recovery 
application. This means evaluating all the recovery actions that constitute the RL 
program. The evaluation of a recovery action is done as follows: each guard is 
translated into a query message for the BB sends back the truth-value of the guard. 
When a guard is found to be true, its corresponding actions are executed, otherwise 
they are skipped.
Ariel is also a configuration language (CFL), that is to say a linguistic framework 
that can be used to reduce to a minimum the code intrusion necessary to include in the 
application a set of fault-tolerance provisions. As an example, the adoption of a 
software watchdog requires the user of the service language to intrude in the code a 
number of non-functional lines of code for the connection, control, and disconnection 
of the watchdog service.
The CFL programmer only sees a high level API with which he can configure a 
specific instance of a fault-tolerance provision, with no need of being aware of even 
which specific software or hardware tool will be used on the target system to 
implement the provision.
The third key attribute of Ariel is that of being a compositional language (CML), 
that is, a linguistic framework with which it is possible to obtain sophisticated 
mechanisms by putting together some ``building blocks''. In this case, Ariel can be 
used as a CML for fast-prototyping what we call a dependable mechanisms (DM) by 
weaving together one or more instances of our fault-tolerance provisions. DM's can 
be defined as high-level software mechanisms that provide a higher dependability 
than the one offered by its building blocks---the fault-tolerance provisions.
Figure~\ref{f:tiran} portrays the TIRAN architecture and its key components.
In the following paragraph, we briefly summarize the specific differences
between RL and other novel approaches.

\SubSection{Specific Differences with respect to Other Approaches}
Numerous techniques have been devised in the past to solve the problem
of optimal and flexible development of dependability services to be embedded
in the application layer of a computer program.
In~\cite{DF00}, some of these approaches are critically reviewed and qualitatively assessed
with respect to a set of structural attributes (separation of design concerns,
syntactical adequacy and adaptability).
A non-exhaustive list of the systems and projects implementing 
these approaches is also given in the cited reference.
In particular, approaches based on metaobject protocols~\cite{KirB91} (MOPs),
FT distributed programming languages~\cite{Rob99}
and aspect-oriented programming~\cite{KLM97} (AOP)
are reviewed therein. In the following, we briefly focus on
MOPs and AOP, two of the most effective and elegant structuring techniques for
flexible development of services.
The main differences between these approaches and RL are remarked. 

\SubSubSection{Metaobject Protocols}\label{ss:mops}
The key idea behind MOPs is that of ``opening'' the implementation of the
run-time executive of an object-oriented language like C++ or Java so that
the developer can adopt and program different, custom semantics,
adjusting the language to the needs of the user and to the requirements
of the environment.
Using MOPs, the programmer can modify the behavior of fundamental features like
methods invocation, object creation and destruction, and
member access. 
The key concept behind MOPs is that of \emph{computational reflection}, or the
causal connection between a system and a meta-level description
representing structural and computational aspects of that system~\cite{Maes87}.
An architecture supporting this approach is FRIENDS~\cite{FaPe98}.
FRIENDS implemented a number of FT provisions (e.g., replication, group-based
communication, synchronization, voting) as MOPs.

A number of studies confirm that MOPs
reach efficiency in some cases~\cite{KirB91}, though no experimental
or analytical evidence allows to estimate the practicality and
the applicability of this approach~\cite{RaXu95,LiLo00}.
MOPs only support object-oriented programming
languages and require special extensions or custom programming languages.
\SubSubSection{Aspect-oriented Programming Languages}\label{ss:aop}
Aspect-oriented programming~\cite{KLM97} is a
programming methodology and a structuring technique
that explicitly addresses,
at system-wide level, the problem of the best code structure
to express different, possibly conflicting design goals
like for instance high performance, optimal memory usage, or
dependability.

Developed as a Xerox PARC
project, AspectJ is an aspect-oriented
extension to the Java programming language~\cite{Kic00,LiLo00}.
A study has been carried out on the
capability of AspectJ as an AOP language supporting
exception detection and handling~\cite{LiLo00}. It has been
shown how AspectJ can be used to develop so-called
``plug-and-play'' exception handlers: libraries of
exception handlers that can be plugged into many
different applications. This translates into better
support for managing different configurations
at compile-time. Up to now, no AOP tool or programming
language exists for flexible development of
{\em dependable\/} services:
AspectJ only addresses exception detection
   and handling. Remarkably enough, the authors of
   a recent study on AspectJ and its support to this field
   conclude~\cite{LiLo00} that ``whether the properties of
   AspectJ [documented in this paper] lead to programs
   with fewer implementation errors and that can be changed easier,
   is still an open research topic that will require serious
   usability studies as AOP matures''.

\Section{The Redundant Watchdog in Ariel}
In order to achieve the Redundant Watchdog functionality described in Section 2 the 
the full linguistic support (CFL, CML, RL) provided by Ariel has been exploited to 
allow the following elements from the TIRAN architecture to work together: 1) the 
RTOS API, and specifically its function TIRAN\_Send, which multicasts a message 
to a logical, 2) the Watchdog, i.e., a node-local error detection provision, and 3) the 
Backbone and its database.
The following scenario is assumed: a distributed system consisting of at least three 
nodes N1, N2, and N3, and on each node of this system, an instance of the TIRAN 
watchdog is running. On a fourth node, N4, or on one of the three watchdog nodes if 
just three nodes are available, an application task is running.
First of all a configuration step is needed in order to:
\begin{itemize}
\item Define and configure the user application tasks
\item Define and configure the Backbone
\item Define and configure the three watchdogs, in particular to assign them the unique-ids W1, W2 and W3
\item Deploy the watchdogs on different nodes and to state that, on a missed deadline, a 
notification is to be sent to the Backbone.
\end{itemize}

Such a configuration step is coded in the Ariel CFL as follows:

\begin{small}
\begin{verbatim}
INCLUDE "watchdogs.h"
  TASK 1 = "Backbone0" IS NODE {N1},
     TASKID {BACKBONE_TASKID}
  TASK 2 = "Backbone1" IS NODE {N2},
     TASKID {BACKBONE_TASKID}
  TASK 3 = "Backbone2" IS NODE {N3},
     TASKID {BACKBONE_TASKID}
  TASK {CLIENT} IS   NODE {N1}, TASKID {CLIENT}
  TASK {W1} IS   NODE {N1}, TASKID {W1}
  TASK {W2} IS   NODE {N2}, TASKID {W2}
  TASK {W3} IS NODE {N3}, TASKID {W3}
  WATCHDOG {W1} WATCHES {CLIENT}
   HEARTBEATS EVERY {BEATCOUNT} MS 
   ON ERROR WARN BACKBONE
  END WATCHDOG
  WATCHDOG {W2} WATCHES {CLIENT}
   HEARTBEATS EVERY {BEATCOUNT} MS 
   ON ERROR WARN BACKBONE
  END WATCHDOG
  WATCHDOG {W3} WATCHES {CLIENT}
   HEARTBEATS EVERY {BEATCOUNT} MS 
   ON ERROR WARN BACKBONE
  END WATCHDOG
\end{verbatim}
\end{small}

The corresponding output is a source file for instantiating three watchdog tasks, 
identified within the user application context as tasks W1, W2, W3, watching 
correspondent application tasks, with a heartbeat rate of \texttt{BEATCOUNT} milliseconds 
and with the default action of sending a warning message to the backbone task when a 
heartbeat is missing. From the user viewpoint, the only code to be intruded in the 
source code of the watched application tasks is given by the macro \texttt{HEARTBEAT}, 
which is translated into the commands for sending a heartbeat message to the 
watchdog tasks. 
Note also that the actual location of the watchdogs is fully transparent to the 
application tasks, as the introduction of these details is done in a separate 
environment, i.e., the configuration program.
A composability step is then required to define tasks W1, W2 and W3 as the logical 
L. This is coded in the Ariel CML as follows:

\begin{verbatim}
 LOGICAL {L} IS TASK {W1}, TASK {W2}, TASK {W3}
 END LOGICAL
\end{verbatim}

Finally there is a recovery step. When a watched task sends ``watchdog L'' its 
heartbeats, the TIRAN\_Send function relays these messages to the three watchdogs 
on the three nodes. In absence of faults, the three watchdogs process these messages 
in the same way---each of them in particular resets the internal timer corresponding to 
the client task that sent the heartbeat. When a heartbeat does not reach a watchdog, 
the watchdog timeout will expire, and the watchdog sends a notification to the BB 
that reacts by wakening the interpreter of Ariel and the r-codes are interpreted. With 
different Ariel code we can easily implement different recovery strategies, and three 
of them have been prototyped in TIRAN:
\begin{itemize}
\item an ``AND-strategy'', that triggers an alarm when each and every watchdog notifies 
BB, 
\item an ``OR-strategy'', the alarm is triggered when any of the three watchdog expires, 
\item a ``2-out-of-3 strategy'', in which a majority of the watchdogs needs to notify 
BB in order to trigger the alarm. 
\end{itemize}

Let us discuss first the AND-strategy, those Ariel code is shown below:
\begin{verbatim}
 IF [ PHASE (TASK{W1}) == {EXPIRED} AND
   PHASE (TASK{W2}) == {EXPIRED} AND
   PHASE (TASK{W3}) == {EXPIRED} ]
 THEN
   SEND {ALARM} TASK{A}
   REMOVE PHASE LOGICAL {L} FROM ERRORLIST
 FI
\end{verbatim}
The guard \texttt{PHASE(TASK\{W\}}\emph{j}\texttt{\})} refers to the info stored by the BB in
its database: upon each alarm received by BB from Wj the corresponding phase is set to 
``expired''. Therefore  the guard evaluates to true only when all three watchdogs have 
expired. The action taken is to reset the phase for the tasks of logical L (action 
REMOVE)and to send an alarm (in the current prototype, the alarm from the 
redundant watchdog is a notification to the task the global identifier of which is A).
The OR strategy can be obtained by changing the
AND operators into OR, and the 2-out-of-3 simply requires to count if at least two watchdogs are in phase EXPIRED.
The different properties for the three strategies can then summarized as follows 
(where the number in parenthesis will be used later during the analysis): 
\begin{itemize}
\item The OR-strategy triggers the alarm as soon as any of the watchdog expires (o1). 
This tolerates the case in which up to two watchdogs have crashed, or are faulty, 
or are unreachable (o2). This intuitively reduces the probability that missing 
heartbeat goes undetected hence can be regarded as an ``integrity-first'' strategy 
(o3). At the same time, the probability of ``false alarms'' (mistakenly triggered 
alarms) is increased (o4). Such alarms possibly lead to temporary pauses of the 
overall system service (o5), with possible implications on the service costs.
\item The AND-strategy, on the other hand, requires that all the watchdogs reach 
consensus before triggering the system alarm (a1). It does not tolerate a crash of 
even a single watchdog (a2). It decreases the probability of false alarms (a3) but 
at the same time decreases the error detection coverage of the watchdog BT. It 
may be regarded as an ``availability-first'' strategy. Should be less expensive than 
OR policy (a4).
\item Strategy 2-out-of-3 requires that a majority of watchdogs expire before the 
system alarm is executed. Intuitively, this corresponds to a trade-off between the 
two above strategies.
\end{itemize}
\Section{Modelling}
In this section we describe how modelling can be used to compare the different 
policies that can be defined using Ariel. Due to lack of space we concentrate only on 
the AND and OR policy (the 2-out-of-3 being an intermediate case), and we only 
show a few results, but the process needed for producing additional ones should be 
clear enough by the end of the section.
From the performance and dependability point of view the most interesting part of the 
RWD is that concerning the alarm and the distinction between the alarm having 
expired because of a real failure of the application (the application is in a halt state) or 
because of delays (either in the application or in the communication network), called 
false alarms. Moreover we have, of course, to consider the possibility of a fault in any 
watchdog. The abstraction level chosen for the analysis assumes that:
\begin{itemize}
\item an application is either working or faulty
\item a watchdog is either working or faulty
\item an application can get out of a faulty state only if the watchdog expires
\item a watchdog can expire due to the fact that, the application heartbeat is not received 
in due time, or the application controlled by the watchdog is faulty, or the 
communication link is broken.
\end{itemize}
Additional assumptions that have been made are about communications for which no 
explicit model is provided, but the same hypothesis are used in all policies, as will be 
explained in the next paragraphs.
The modelling formalism used is that of Generalized Stochastic Petri Nets (GSPN) 
\cite{GSPN} in which transitions are either immediate (and they fire in zero time) or 
timed (with an associated exponentially distributed delay). The tool
GreatSPN~\cite{CFGR96} has been used which allows the computation of performance measures 
using either steady state or transient analysis, as well as simulation (that allows also 
the solution of models with transitions that have generally distributed delays, as 
deterministic, gaussian, etc.)
Figure 2 shows the GSPN model of a redundant watchdog made up of three 
watchdogs with OR policy.  The model is composed of a skeleton application (left 
portion) and a skeleton watchdog (right portion). Places starting with Ap are part of 
the application model, while places starting with Wd are part of the watchdog model.
Both application and watchdogs can be faulty, but let us describe first the normal 
behaviour. The application performs a computation (transition activity) and then sends 
a kick to the watchdog (actually to the logical that is composed of three watchdog 
processes), modelled by transition ok. The redundancy level of the watchdog is 
realised by assigning an initial marking equal to 3 to place Wd1. In the watchdog, if a 
heartbeat message arrives before the timeout expires (that is to say before the firing of 
transition timeout), transition ok will fire, removing 3 tokens from Wd1; if instead one 
of the timeouts expires before the kick arrives, since we are modelling the OR policy, 
transition delayed will fire, to model the case of an application that is not faulty, but 
simply too slow with respect to the chosen value for the timeout. Observe that this 
takes substantially into account also the case of an application that sends the heartbeat 
message in time, but the message gets delayed in the network.
The application, or one or more of the watchdogs, can go into an halting state due to 
an error caused by a fault. This is modelled by transition ap-fault for the application 
and by transition w-fault for the watchdogs. As a consequence, it is now possible that 
the timeout expires because the application is in an halting state (place Ap2), and this 
is modelled by transition faulty. Another possible scenario is that one or more 
watchdogs go into a halting state. In that case the remaining non-faulty watchdogs 
still perform their count-down, unless all of them are faulty (three tokens in place 
Wd3). 
The GSPN model of a redundant watchdog made with AND policy differs only in 
small, but significant, details: the multiplicity of the arcs from Wd2 to transitions 
delayed and faulty is  fixed to three, since all three timeouts should expire before an 
action with respect to the application is taken. Moreover it is possible that, when a 
heartbeat arrives, some of the timeouts have already expired, so that also place Wd2 is 
an input place for transition ok.

\begin{figure*}[tpb]
\centerline{\includegraphics[width=.7\textwidth]{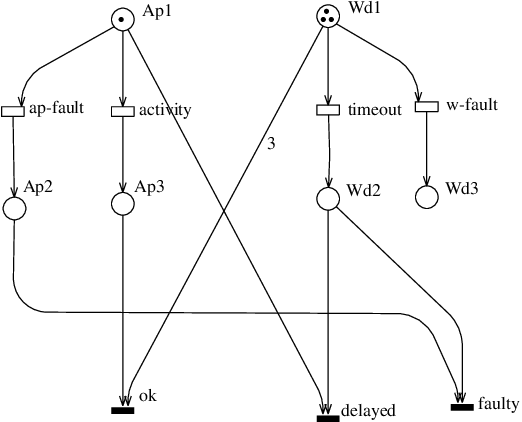}}
\caption{The redundant watchdog with OR policy.}
\end{figure*}

The model that has been used for the solution differs from the one depicted above 
since we want to consider a cyclic behaviour, as this is typical of the automation 
environment addressed by this work, that is to say an application that performs an 
activity and sends the heartbeat message in an endless loop. A number of arcs and 
transitions have been added to the model to reset back to their initial states the 
application and the watchdogs, and an additional delay has been inserted for this re-
cycling activity (transition cycle): this results in an ergodic model that can be solved 
in steady state.
Weights have been assigned in a rather blind (non realistic) manner, since the goal of 
this preliminary analysis is to compare the different policies, and not to produce 
absolute measures. Transition activity has a mean delay of 0.5, w-fault and ap-fault a 
delay 20 times bigger, and the recycling activity is set to 1.0. The rate of transition 
timeout has been taken as a varying parameter from 0.5 to 2.0 (for corrisponding 
delays varying from 2.0 to 0.5) 
To decide which performance measures to use for the comparison, we can consider 
the properties o1-- o5 of the OR strategy and a1--a4 of the AND strategy listed at the 
end of the previous section, and for each property we identify a corresponding 
measure to be computed or a property to be proved.
\begin{description}
\item[(o1)] place Wd2 is always empty 
\item[(o2)] there is a state in which delayed o faulty can fire, although Wd3 is  $\ge 2$
\item[(o3)] throughput of transitions delayed and faulty
\item[(o4)] throughput of transition delayed
\item[(o5)] throughput of transition activity (useful work)
\end{description}
\begin{description}
\item[(a1)] structural property due to the weight 3 on the arc from Wd2 to transitions 
delayed and faulty, and we can also check that there is a non null probability of 
having 2 tokens in place Wd2.
\item[(a2)] existence of a P-invariant stating that the sum of the tokens in Wd places is equal 
to 3, therefore since a fault in a watchdog puts a token in Wd3, then Wd2 will never 
have more then 2 tokens, and therefore delayed and fault will never fire.
\item[(a3)] throughput of transition delayed 
\item[(a4)] throughput of transition activity (useful work)
\end{description}
Properties (o1), (o2) have been proven by inspecting the state space (that contains 
only a few dozens states), (a2) has required a P-invariant computation, while all other 
properties are based on a comparison of the throughputs of the transitions activity, 
delayed and faulty for the two models, for varying values of the delay associated with 
the timeout transition, that are reported in the diagram of Figure 3 var varying values 
of the rate of transition timeout (note that the throughput of transition faulty has not 
been shown since too small for the given choice of  parameters, and that we have also 
reported the throughput of transitions cycle and timeout) 
The throughput of activity shows how much work is actually performed by the 
application, and it is clearly greater for the AND policy. Throughput of cycle shows 
that the OR policy causes more restarts than the AND one, while the throughput of 
delayed (marked del in the legenda) shows that OR sends more false alarms to the 
backbone than the AND policy. Finally, the higher throughput of timeout for the AND 
policy is due to the fact that the AND policy has less restart of the watchdogs than the 
OR one. Additional analysis is instead needed to show that the OR policy has a lower 
``time to detect an application fault''.

\begin{figure*}[tpb]
\centerline{\includegraphics[width=.9\textwidth]{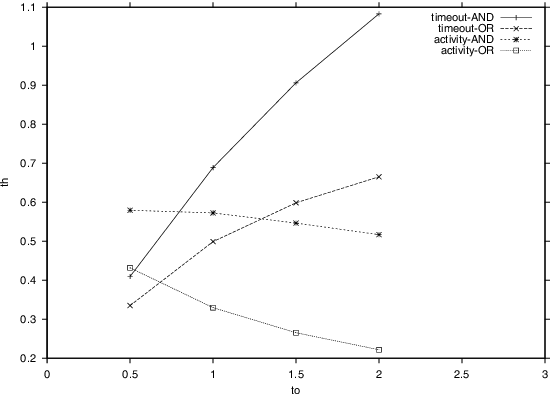}}
\caption{Comparison of AND and OR policies.}
\end{figure*}

\Section{Conclusions and Further Developments}
A novel approach for the development of dependable and flexible automation services 
has been introduced in the project TIRAN and it has been illustrated here by mean of 
the redundant watchdog. The approach is based on the compositional capabilities of 
Ariel, a custom language for error recovery and configuration, and it allows to 
develop dependable tools whose flexibility allows the user to easily set up automation 
services fulfilling very different dependability requirements. This flexibility gives 
more freedom to the dependability designer, that now has to face the problem of 
comparing them: in our case study we have used stochastic modelling based on Petri 
net to achieve such a comparison.
Summarising, the concept of recovery language allows to express the application 
software as two separate codes: the functional code and the r-code. The former deals 
with the specification of the functional service, whereas the latter is the description of 
the measures that need to be taken in order to perform some corrective actions, such 
as ordering the modification of some key parameter like, for instance, the code 
redundancy used in data transmission, or which software process needs to be 
appointed to a given sub-task. The specification of these corrective actions is done by 
the user in an environment other than the one for the specification of the functional 
aspects. Furthermore, this separation still holds at run-time, since the executable code 
and the r-code are physically distinct. This strict separation between the two aspects 
may allow to ``trade'' at run-time the actual set of recovery actions to be 
executed---which may be exploited, for instance, to provide a mobile code with the 
required adaptability to different environment conditions.
The reported approach is currently being further developed and experimented within 
the recently started IST Project 25434 DepAuDE (Dependability for embedded 
Automation systems in Dynamic Environments with intra-site and inter-site 
distribution aspects).

\bibliographystyle{latex8}
\bibliography{./thesis}
\end{document}